\documentclass[conference]{IEEEtran}
\IEEEoverridecommandlockouts
\usepackage{cite}
\usepackage{amsmath,amssymb,amsfonts}
\usepackage{graphicx}
\usepackage{textcomp}
\usepackage{xcolor}
\usepackage{bbm}
\usepackage{amsthm}
\theoremstyle{definition}
\usepackage{algpseudocode}
\usepackage{algorithm}
\usepackage{float}
\usepackage{hyperref}
\usepackage{listings}
\usepackage[normalem]{ulem}
\def\BibTeX{{\rm B\kern-.05em{\sc i\kern-.025em b}\kern-.08em
    T\kern-.1667em\lower.7ex\hbox{E}\kern-.125emX}}
\begin{document}

\newcommand{\setsuccess}[1]{S_{#1}^{\mathsf{s}}}
\newcommand{\setfail}[1]{S_{#1}^{\mathsf{f}}}
\newcommand{\veri}{\operatorname{ver}_{f, \varphi}}
\newcommand{\simu}{\operatorname{task}_{f, \varphi}}
\newcommand{\rob}{\operatorname{rob}_{f, \varphi}}
\newcommand{\ourproblemfull}{Repair with Preservation (RwP)}
\newcommand{\ourproblem}{RwP}
\newcommand{\ourmethodfull}[0]{Incremental Simulated Annealing Repair (ISAR)}
\newcommand{\ourmethod}{ISAR}
\newcommand{\notimplies}{\mathrel{{\ooalign{\hidewidth$\not\phantom{=}$\hidewidth\cr$\implies$}}}}

\newtheorem{num1}{Theorem}
\newtheorem{num2}{Lemma}
\newtheorem{num3}{Remark}
\newtheorem{num4}{Definition}
\newtheorem{num5}{Assumption}
\newtheorem{num6}{Example}
\newtheorem{theorem}[num1]{Theorem}
\newtheorem{lemma}[num2]{Lemma}
\newtheorem{remark}[num3]{Remark}
\newtheorem{definition}[num4]{Definition}
\newtheorem{assumption}[num5]{Assumption}
\newtheorem{example}[num6]{Example}
\newtheorem{proposition}[num1]{Proposition}
\newtheorem{corollary}[num1]{Corollary}

\newcommand{\EL}[1]{{\color{purple}#1}}
\newcommand{\IR}[1]{{\color{blue}#1}}
\newcommand{\MC}[1]{{\color{brown}#1}}
\newcommand{\OS}[1]{{\color{blue}#1}}
\newcommand{\IL}[1]{{\color{red}#1}}

\newcommand{\edit}[1]{#1}

\title{Repairing Learning-Enabled Controllers\\While Preserving What Works}
% \title{Repair Only What is Broken\\in Learning-enabled Controllers}
% \title{Correctness Preserved: Global Repair of Learning-enabled Controllers without Compromising Correct Scenarios}
% \title{Consistently Correct: Global Repair of Learning-enabled Controllers without Compromising Correct Scenarios}

% Some candidate titles by Chat GPT:
% 1. Safeguarding Repaired Inputs: A Novel Algorithm for Sustained Global Repair in Learning-enabled Controllers
% 2. Resilient Global Repair: A Progressive Algorithm for Robust Learning-enabled Controller Input Corrections
% 3. Unyielding Global Repair: An Algorithm that Upholds Repaired Learning-enabled Controller Inputs Amidst Ongoing Corrections
% 4. Consistently Correct: A Sustained Approach to Learning-enabled Controller Global Repair without Compromising Repaired Inputs
% 5. Guardians of Repair: Ensuring the Integrity of Corrected Inputs in Global Repair of Learning-enabled Controllers

% For aligning authors
\makeatletter
\newcommand{\linebreakand}{%
  \end{@IEEEauthorhalign}
  \hfill\mbox{}\par
\mbox{}\hfill
\begin{@IEEEauthorhalign}
}

\author{
\IEEEauthorblockN{Pengyuan Lu}
\IEEEauthorblockA{\textit{Computer and Information Science}\\ \textit{University of Pennsylvania}\\
Philadelphia, PA, USA\\
pelu@seas.upenn.edu}
\and
\IEEEauthorblockN{Matthew Cleaveland}
\IEEEauthorblockA{\textit{Electrical and Systems Engineering}\\ \textit{University of Pennsylvania}\\
Philadelphia, PA, USA\\
mcleav@seas.upenn.edu}
\and
\IEEEauthorblockN{Oleg Sokolsky}
\IEEEauthorblockA{\textit{Computer and Information Science}\\ \textit{University of Pennsylvania}\\
Philadelphia, PA, USA\\
sokolsky@seas.upenn.edu}
\linebreakand
\IEEEauthorblockN{Insup Lee}
\IEEEauthorblockA{\textit{Computer and Information Science}\\ \textit{University of Pennsylvania}\\
Philadelphia, PA, USA\\
lee@seas.upenn.edu}
\and
\IEEEauthorblockN{Ivan Ruchkin}
\IEEEauthorblockA{\textit{Electrical and Computer Engineering}\\
\textit{University of Florida}\\
Gainesville, FL, USA\\
iruchkin@ece.ufl.edu}
}
% \and
% \and
% \and
% \and
% \and

\maketitle

\begin{abstract}
Learning-enabled controllers have been adopted in various cyber-physical systems (CPS). When a learning-enabled controller fails to accomplish its task from a set of initial states, researchers leverage repair algorithms to fine-tune the controller's parameters. However, existing repair techniques do not preserve previously correct behaviors. Specifically, when modifying the parameters to repair trajectories from a subset of initial states, another subset may be compromised. Therefore, the repair may break previously correct scenarios, introducing new risks that may not be accounted for. Due to this issue, repairing the entire initial state space may be hard or even infeasible. As a response, we formulate the Repair with Preservation (RwP) problem, which calls for preserving the already-correct scenarios during repair. To tackle this problem, we design the Incremental Simulated Annealing Repair (ISAR) algorithm, which leverages simulated annealing on a barriered energy function to safeguard the already-correct initial states while repairing as many additional ones as possible. Moreover, formal verification is utilized to guarantee the repair results. Case studies on an Unmanned Underwater Vehicle (UUV) and OpenAI Gym Mountain Car (MC) show that ISAR not only preserves correct behaviors from previously verified initial state regions, but also repairs 81.4\% and 23.5\% of broken state spaces in the two benchmarks. Moreover, the average \edit{signal temporal logic (STL)} robustnesses of the ISAR repaired controllers are larger than those of the controllers repaired using baseline methods.
\end{abstract}

\begin{IEEEkeywords}
controller repair, neural network repair,  learning-enabled controller
\end{IEEEkeywords}
\section{Introduction}
\label{sec:intro}

% Overview of what we are doing
Learning-enabled controllers are increasingly being adopted in various cyber-physical systems (CPS), including autonomous vehicles, industrial control, and healthcare services \cite{yu2021reinforcement,escandell2014optimization,mosquera2023enabling,tuncali2018reasoning}. 
When a learned controller, such as a deep neural network, fails to accomplish its pre-defined task in some scenario, its engineers or users would like to fix its performance ideally without sacrificing \edit{the performance in other scenarios.} Researchers have proposed various repair techniques to adjust a neural network's parameters,
so that the network is repaired on a given set of inputs \cite{fu2022reglo,zhou2020runtime}.
% so that the network is repaired on multiple inputs instead of a single input.
% % so that the network performs well on multiple inputs rather than just one.
% % \OS{Consider restating this sentence.  Unclear what "just one" means here.} 
% These techniques are referred to as \textit{global repair}~\cite{fu2022reglo}.

% When a learned controller, such as a deep neural network, fails to accomplish its pre-defined tasks, we can apply repair techniques to fix it. Researchers in the repair domain have proposed various algorithms to adjust a neural network's parameters, so that the network can achieve a given goal on a set of multiple inputs. However, state-of-the-art techniques have not yet addressed the competing nature among the inputs, that tuning the parameters to fulfill the goal on one input may compromise the performance on another. Therefore, global repair boils down to a multi-objective optimization problem \cite{sawaragi1985theory,zitzler2004tutorial,collette2004multiobjective}, 
% with repairing the system on each input being a distinct objective. It would be promising if we can identify appropriate Pareto-optimal solutions on the objectives trade-off, such as a controller that repairs on as many inputs as possible under the given dynamics and the capacity of neural network.

% and it would be promising if we can identify appropriate Pareto-optimal solutions for specific learning-enabled controller applications. 

% Motivating example - patient-caring robot from multiple initial locations
Consider a motivating example of a healthcare service robot \cite{wan2020cognitive,holland2021service}. The robot needs to attend to the patient in bed within a given time limit after an alarm is raised. Before the alarm, the robot can be in various physical locations, such as \edit{the living room or its charging station}. Now, upon the alarm, the robot is able to reach the bed in a timely manner from almost all locations, except for the set of initial states in kitchen, where it loiters around and wastes time. 
Suppose the engineers have identified that the problem stems from the neural network-based controller. Then 
repair algorithms can be leveraged ~\cite{fu2022reglo,zhou2020runtime} to fix the problem.
% this scenario leads to a global repair problem
% ~\cite{fu2022reglo,zhou2020runtime}.\OS{is this description of global repair consistent with the one in the previous paragraph?  In any case, do we need both?} 
Specifically, the ultimate goal is to adjust the controller network so that the system accomplishes the task from all initial locations, including the kitchen and other previously successful locations.

% Specifically, the objective is to maintain the performance of the robot from all the already-working initial states, while incrementing as many new working initial states (from the kitchen corner area) as possible.

% Here, $T$ is the time limit, or deadline, that the robot needs to reach the bed within. The real-valued function $\rho$ measures how well a trajectory of states $s_0^{(i)}, \dots, s_T^{(i)}$ accomplishes the task. This measurement function can be obtained from off-the-shelf techniques such as quantitative semantics score \cite{hamilton2022training}. From this formulation, we can see the potential competition among the trajectories -- a solution $\pi'$ may improve the score $\rho$ on some trajectories but compromise the others. Since the system dynamics $f$ is deterministic, the competition among trajectories is in fact a competition among the initial states in $S_{init}$. Therefore, a solution $\pi'$ can hardly or infeasible to be optimal for all initial states \cite{collette2004multiobjective}. Instead, we seek a Pareto-optimal $\pi'$ on the trade-off among these initial states.

However, state-of-the-art techniques \cite{zhou2020runtime,fu2022reglo,santa2022safe,yang2022neural} 
\edit{may not preserve} previously successful behaviors during repair. Specifically, previously successful behaviors may be broken after the controller parameters are altered. This is because of the competing nature among the initial states: When the repair fixes the trajectories from a subset of initial states, the performance of trajectories from another subset may be compromised. In the healthcare robot example, a repair algorithm may correct the trajectories from the kitchen but break the trajectories from elsewhere, such as the living room. This may lead to unseen risks such as the robot tripping or crashing into objects, posing new potential dangers to the patient. Generally, it may be hard or even infeasible to identify a controller to accomplish the task from all initial states. Due to this reason, existing literature that aims to  repair controller neural networks on the entire initial state space cannot establish guaranteed outcomes~\cite{zhou2020runtime,santa2022safe,yang2022neural}. By contrast, instead of looking for a controller that correctly behaves under all scenarios, our approach is to preserve what is already correct while repairing as many other scenarios as possible. In the worst case, the repaired controller will still operate correctly on the previously successful initial states. So the repair will not introduce any new erroneous behaviors.

We propose to solve the controller repair problem while safeguarding the previously successful behaviors. Specifically, our ultimate goal is to find an alternative controller $\pi'$ that (i) still accomplishes the task from previously successful initial states and (ii) maximizes the number of additional repaired initial states. We denote this problem as the \ourproblemfull{} problem.
% As guaranteeing the maximality of repaired initial states is hard, we seek a lower-hanging fruit. That is, we aim to develop an algorithm that incrementally increases the set of repaired initial states while guaranteeing the protection of the previously working initial states. 
We identify three challenges when tackling this problem. First, the set of initial states may be uncountably infinite, posing difficulty in checking the performance from all initial states. Second, we need to protect the performance on the initial states for which the controller behaves correctly while repairing the performance of the controller on the other initial states. Third, we \edit{would like to} formally prove that our repaired controller network works on both the previously successful initial states, as well as the newly repaired ones. 

To solve this problem, we develop the \ourmethodfull{} algorithm, which safeguards the previously successful initial states during repair and returns verified outcomes afterwards. To use this algorithm, the initial state set is partitioned into finitely many regions. Each region first passes through a sound but incomplete verifier, such as Verisig~\cite{ivanov2019verisig,ivanov2021verisig}, to formally prove whether trajectories from this entire region are able to accomplish the task. During repair, simulated annealing~\cite{kirkpatrick1983optimization,serafini1994simulated,suman2006survey} 
\edit{is applied to a barrier-guarded energy function\cite{polyak1992modified,den1992classical,hauser2006barrier}}
% Monte Carlo-integrated\MC{This is a property of our algorithm, not of the function itself}~\cite{robert1999monte} energy function 
to greedily repair the initial state regions one-by-one, while protecting the previously successful regions. After repair, the initial state regions will be checked by the verifier again to produce formally proven outcomes.

Case studies on an Unmanned Underwater Vehicle (UUV) and Mountain Car (MC) \cite{brockman2016openai} demonstrate that \ourmethod{} is able to preserve verified correct behaviors \edit{specified in signal temporal logic (STL)}. Specifically, initial state regions that pass verification before repair can consistently be verified afterwards. Moreover, $81.4\%$ and $23.5\%$ of regions that fail verification can be repaired on UUV and MC, respectively. We show that \ourmethod{} defeats baseline methods in STL robustness, which measures STL-specified task performance. \edit{Our implementation can be found at \texttt{\url{https://github.com/ericlupy/isar_rep}}, with repeatability details in the Appendix.}
% The overall STL robustness also defeat baseline methods.\IL{I don't understand this setnece.  STL robustness is mentioned out of blue.}

% Contributions list
Overall, we make the following contributions.
\begin{enumerate}
    \item We formulate the \ourproblemfull{} problem, which calls for protecting the initial states that already produces task-accomplishing trajectories during repair. %\IL{Are these repairs applied during the design time or after the system has been deployed?}
    \item We propose the \ourmethodfull{} algorithm to tackle the \ourproblem{} problem, % \MC{not sure why the spacing is weird here. IL fixed it :)}
     which safeguards the already-correct initial states and produces assurance on repair results via formal verification. This repair algorithm runs offline after task failures are identified in a deployed system.
    \item We run \ourmethod{} on two case studies: an unmanned underwater vehicle (UUV) and OpenAI Gym Mountain Car (MC). Results show that \ourmethod{} is able to preserve correctness on verified initial state regions, while repairing 81.4\% and 23.5\% of broken state spaces, respectively. The \ourmethod{}-repaired controller also demonstrates higher STL robustness than baselines.
\end{enumerate}

\section{Related Work}
\label{sec:related_work}

Neural network repair algorithms can be divided into two major categories: modifying outputs and modifying network parameters. \edit{For modifying outputs, previous works generally patch the controller outputs to keep the system} within a safe bound \cite{lyu2023autorepair,sohn2019search,tokui2022neurecover,lu2023causal}. However, since these methods modify on the outputs instead of the network parameters, the errors still persist if the same network is reused.
% However, this approach does not produce reusable models to accomplish pre-defined tasks.\IL{What does this mean? And, then why?}

% Re-written - local vs. global are not standard terms
Researchers have also developed repair techniques that modify neural network parameters. Some consider specific types of networks, such as ReLU-activated layers \cite{fu2021sound} and two-level lattice structures \cite{santa2022safe}. Some focus on repairing on a finite set of inputs \cite{sotoudeh2021provable} or local input neighborhoods \cite{majd2021local}, \edit{and these procedures usually exhibit provable soundness, due to the assumption of being finite \cite{fu2022reglo}.}
% where soundness can be relatively easily proven compared to repairing on an entire continuous input space \cite{fu2022reglo}\MC{I don't quite follow this sentence}. 
Different techniques are also adopted, including satisfiability modulo theory (SMT) solvers \cite{cohen2022automated}, causality analysis \cite{sun2022causality} and formal methods \cite{dong2020towards,usman2021nn}. The majority of repair publications work on general-purpose neural networks, while some literature concerns neural network-enabled controllers \cite{zhou2020runtime,santa2022safe,yang2022neural} that can be leveraged in CPS --- here, the controller is repaired to produce state trajectories that accomplish a task, such as staying within a safe region.

% When modifying neural network parameters, scientists further divide the problem into two sub-categories: local and global repair. In local repair, the algorithm aims to produce a task-accomplishing output on a specific input, optionally including its surrounding neighborhood. Correctness of local repair methods are relatively easy to be proven, with the algorithm leveraging satisfiability modulo theory (SMT) solvers \cite{cohen2022automated}, causality analysis \cite{sun2022causality}, verification \cite{dong2020towards} or other correct-by-construction techniques \cite{fu2021sound,sotoudeh2021provable,majd2021local,majd2023safe}. Global repair aims to produce task-accomplishing outcomes on a (usually continuous) set of inputs. Correctness of global repair methods are generally harder to be proven, except that proofs exist for particular repair goals such as global robustness \cite{fu2022reglo}. So far, researchers have discussed global repair on neural networks for general-purposes \cite{fu2022reglo,usman2021nn}, as well as neural network-based controllers that can be leveraged in CPS \cite{zhou2020runtime,santa2022safe,yang2022neural}. 
% % The latter aims for repairing a controller to satisfy a given property from a set of initial states, matching our second sub-problem.\IL{What does it? Why is your point here?}

So far, the repair literature has yet to discuss preserving correct behaviors during the repair. We are aware that researchers in machine learning (continual and transfer learning \cite{chen2018lifelong,liu2017lifelong,silver2013lifelong,ruvolo2013active}, in particular) have already discussed a relevant topic, i.e., catastrophic forgetting prevention \cite{robins1995catastrophic,french1999catastrophic,serra2018overcoming}. The difference between continual learning and repair is that the former adapts knowledge from one explicit domain to another, while the latter focuses on improving performance in a static domain. Whether the repair problem can be \edit{cast} as implicit domain adaptation \edit{is left to} future work.
\section{Background}
\label{sec:background}

Our problem setting is based on the following formalization.
Let $t = 0, 1, 2, \dots$ denote the discrete time steps, $S$ be a continuous physical state space, and $A$ be a continuous action space. \edit{Let $\pi:S \mapsto A$ denote the controller}, i.e. the current action $a_t = \pi(s_t)$. \edit{The controller $\pi$ is parameterized by some $\theta \in$ a fixed parameter space $\Theta$, e.g., the weights and biases for some neural network architecture. We use $\pi_\theta$ to denote such parameterization.} The environment dynamics is $f: S \times A \mapsto S$, i.e., the next state $s_{t+1} = f(s_t, a_t)$. Next, we introduce relevant concepts.

\subsection{Signal Temporal Logic (STL) and Robustness Score}
\label{subsec:stl}

Signal temporal logic (STL) is a logical \edit{formalism} to specify properties of continuous signals, such as trajectories of physical states in a continuous state space. In CPS, STL \edit{is used for both monitoring and verification of task } accomplishment \cite{ivanov2019verisig,ivanov2021verisig,ruchkin2022confidence}. On a trajectory denoted as $\bar{s} := s_0s_1\dots s_T$, $\forall s_t \in S$, the grammar of STL is defined as
\begin{equation}
    \label{eq:stl_grammar}
    \varphi ::= \top \mid g(\bar{s}) < 0 \mid \neg\varphi \mid \varphi_1 \land \varphi_2 \mid \varphi_1 U_{[t_1, t_2]} \varphi_2.
\end{equation}
Here, $\top$ is tautology and $g: S^T \mapsto \mathbb{R}$ is a real-valued function on signals. Temporal operator $U_{[t_1, t_2]}$ denotes the until operator, and $\varphi_1 U_{[t_1, t_2]} \varphi_2$ means $\varphi_2$ must hold at some time $t \in [t_1, t_2]$ and $\varphi_1$ must always hold before $t$. The until operator can be converted into two additional temporal operators (i) eventually/finally operator $F_{[t_1, t_2]}$ and (ii) always/globally operator $G_{[t_1, t_2]}$. \edit{The exact definition of STL semantics can be found in the original paper~\cite{maler2004monitoring}.}% Please refer to the original paper \cite{maler2004monitoring} for details. 

Conventional STL semantics produce a Boolean outcome, indicating whether a given trajectory accomplishes its task. STL is also equipped with quantitative semantics~\cite{fainekos2009robustness,gilpin2020smooth,hamilton2022training}, a robustness score to evaluate the degree of a trajectory satisfying an STL formula $\varphi$. Specifically, the robustness score $\rho: S^T \times \mathbb{N}_{\geq 0} \times \Phi \mapsto \mathbb{R}$ is a real-valued function that evaluates $\rho(\bar{s}, t, \varphi)$ on three inputs: a trajectory $\bar{s}$ of length $T$, a starting time step $t$ of the trajectory 
% \OS{There is no $t$ in the definition of $\rho$}
, and an STL formula $\varphi \in \Phi$. The reader is referred to classic sources~\cite{fainekos2009robustness} for its definition. Based on this definition, $\rho \geq 0$ means that the trajectory starts at time $t$ with length $T$ is able to fulfill the STL formula $\varphi$, i.e., it accomplishes the specified task. Likewise, $\rho < 0$ means it fails. The larger absolute value of $\rho$, the higher degree of accomplishment or failure.

We assume all tasks are given in form of STL specifications, and the following subroutine is available to compute performance on a task $\varphi$ from an initial state $s_0$, under dynamics $f$ and controller $\pi$.

\begin{definition}[Robustness Computing Subroutine]
    \label{def:robustness_computing}
    Given a dynamics $f$ and an STL-specified task $\varphi$, a robustness computing subroutine $\rob: S \times \Pi \mapsto \mathbb{R}$ is a function that computes the STL robustness of a trajectory from an initial state $s_0$ under a  controller $\pi$. That is,
    \begin{equation}
        \rob(s_0, \pi) := \rho(\bar{s}, 0, \varphi)
    \end{equation}
    Here, trajectory $\bar{s} = s_0s_1\cdots s_T$, with $a_t = \pi(s_t)$ and $s_{t+1} = f(s_t, a_t)$ for all $t \in [0, T)$. The set $\Pi$ denotes the space of all controllers.
\end{definition}

In other words, $\rob(s_0, \pi)$ evaluates the degree of task accomplishment from initial state $s_0$ under controller $\pi$.

% With knowledge of the deterministic dynamics $f$, the trajectory from an arbitrary initial state can be computed. If the dynamics is not explicitly known, but we can run the CPS's model in the black-box environment or its simulator, trajectories can be obtained by data collection. When a trajectory can be acquired from an arbitrary initial state, we are able to compute its Boolean and quantitative outcomes on an STL-specified task, i.e., a task checking and a robustness computing subroutine are available. Formally, we assume the follows.
% \begin{assumption}[Available Subroutines]
%     \label{assum:available_subroutines}
%     Under dynamics $f$ and task $\varphi$, we have a task checking subroutine $\simu: \mathcal{S} \times \Pi \mapsto \{0, 1\}$ and a robustness computing subroutine $\rob: \mathcal{S} \times \Pi \mapsto \mathbb{R}$.
% \end{assumption}

\subsection{Simulated Annealing}
\label{subsec:sim_annealing}

Simulated annealing is a stochastic optimization algorithm that aims to find global optima of an objective function, \edit{which is often equivalently referred as the \textit{energy function} }
%also known as the energy function 
\cite{kirkpatrick1983optimization,serafini1994simulated,suman2006survey}. We denote the energy
%\MC{Is this the same as the objective function?} 
function as $e: \Theta \mapsto \mathbb{R}$, i.e., a real-valued function that depends on variables $\theta \in \Theta$.

The algorithm works as follows. At every iteration, it randomly perturbs previous variables $\theta$ to obtain new variables $\theta'$, e.g., by adding Gaussian noise to each variable. Then, the energy function is evaluated to see if the new \edit{$e(\theta')$} is better than the previous \edit{$e(\theta)$}
% \MC{Right now $e$ denotes both the energy function itself and a specific value of the energy function.}
. Specifically, if \edit{$\Delta_e := e(\theta') - e(\theta) \geq 0$}, the new variables are accepted. Otherwise, if \edit{$\Delta_e < 0$}, a second check is performed by tossing a biased coin with acceptance probability
\edit{
\begin{equation}
    \label{eq:sa_criterion}
    \Pr[\text{acceptance}] := \exp(-\Delta_e / \tau). 
\end{equation}
}
Here, $\tau > 0$ is \edit{called the temperature, with higher values leading to more exploration.} This second criterion is formally known as Metropolis-Hastings criterion, which encourages exploration of new variables, even if the new energy function is slightly worse. At the end of every iteration, the temperature cools down, e.g., by multiplying a cooling factor $\alpha \in (0, 1)$.
% \MC{How much does it cool down? Can we write out an explicit formula?}.
\edit{So the amount of exploration decreases}  throughout the procedure. The algorithm \edit{runs until either the energy function converges or a maximum number of iterations is reached}.

Compared to other optimization algorithms such as gradient descent, simulated annealing is able to escape local optima due to its stochastic nature and encouragement in exploration. We will use a modified version of this technique in our main algorithm.

\subsection{The Learning-enabled Controller Repair Problem}
\label{subsec:global_repair}

% \MC{Is this a new problem that we are defining, or a standard one that we are simply restating? Also, say ``the existing literature would pose this problem as '' feels awkward to me. }
% We now formalize the global repair problem on a real-time closed-loop CPS controller, such as the healthcare service robot example in Section \ref{sec:intro}. 
% Assume the CPS starts from any state in $S_{init} \subset \mathcal{S}$, 
% the existing literature \cite{zhou2020runtime,santa2022safe,yang2022neural} would pose this problem in the following form: \MC{The cost should also contain $\pi'$}

The repair problem on learning-enabled controllers formulated by existing literature \cite{zhou2020runtime,santa2022safe,yang2022neural} takes a form as follows: \edit{for all $s_0 \in S_{init},$}
% \MC{This equation as two separate uses for $s_0$}
\edit{
\begin{equation}
    \label{eq:example}
    \begin{split}
        &\text{minimize}_{\pi'}\,  c(s_0, \pi, \pi')\\
        &\text{subject to } \rob(s_0, \pi') \geq 0.
    \end{split}
\end{equation}
}Here, $c$ is a generic cost function on initial state $s_0$, current controller $\pi$ and repaired controller $\pi'$. For example, in minimally-deviating repair \cite{zhou2020runtime}, the cost is the difference between the trajectory from $s_0$ produced by the repaired controller $\pi'$ and the original controller $\pi$.
% Here, $c$ is a constant (reducing it to a constraint satisfaction problem) or a cost function to be minimized, such as the deviation from $\pi'$ to the original controller $\pi$ as in minimally-deviating repair \cite{zhou2020runtime}.
Recall that a non-negative robustness is equivalent to task being accomplished. \edit{Therefore, the state-of-the-art is aiming for an ambitious goal, which guarantees task accomplishment while minimizing the cost on \textit{all} initial states.}
% Therefore, the constraint requires all initial states $s_0 \in S_{init}$ to accomplish the task under the system dynamics.
% The constraint requires all initial states $s_0 \in S_{init}$ to (i) satisfy the system dynamics and (ii) produce trajectories $s_0, \cdots, s_T$ that accomplishes the task, such as the robot reaching the bed timely.

\edit{Despite being ideal, solving for Equation \eqref{eq:example} on all $s_0 \in S_{init}$ simultaneously can be hard or even infeasible. This is because modifying the controller parameters to optimize for some initial states may compromise the others --- a competing nature among the initial states under the controller capacity and system dynamics. Therefore, we seek a feasible problem setting.}
% However, solving such a problem may be hard or even infeasible, because the constraint requires the CPS to accomplish the task from all initial states, which may be competing under the controller capacity and system dynamics. Therefore, we seek a feasible problem that is guaranteed to have a solution under the trade-offs among the initial states.\MC{What does this last sentence mean?}

\section{Formulating the \ourproblemfull{} Problem}
\label{sec:problem}

\subsection{A Repair Problem that Considers Initial State Competition}
\label{subsec:problem_feasible}

To identify a feasible problem, we first define a partition of the initial states $S_{init}$ under a controller $\pi$.
\begin{definition}[Initial State Partition]
\label{def:partition}
Given dynamics $f: S \times A \mapsto S$,
% and a task
% $\varphi: \mathcal{S}^T \mapsto \{0, 1\}$,
a controller $\pi: S \mapsto A$ partitions the initial states $S_{init} \subset S$ into \textit{successful} and \textit{failed} ones: $S_{init} = \setsuccess{\pi} \cup \setfail{\pi}$. That is,
\begin{equation}
    \label{eq:partition}
    \begin{split}
        &\setsuccess{\pi} := \{s_0 \in S_{init}: \rob(s_0, \pi) \geq 0\} \\
        % &\forall t = 0, \dots, T,a_t = \pi(s_t)
        % \land s_{t+1} = f(s_t, a_{t})\}\\
        &\setfail{\pi} := S_{init} \setminus \setsuccess{\pi}
    \end{split}
\end{equation}
\end{definition}

Definition \ref{def:partition} formalizes a partition of the initial states under a controller $\pi$. In plain words, $\setsuccess{\pi}$ is the set of initial states \edit{from which $\pi$ satisfies $\varphi$} 
% from which the trajectories \MC{Can we say "initial states from which $\pi$ satisfies $\phi$"?} are able to accomplish the task with $\pi$ 
(``$\mathsf{s}$" stands for ``success"). Likewise, $\setfail{\pi}$ is the set of initial states \edit{from which $\pi$ does not satisfy, or equivalently, fails on $\varphi$} 
% whose trajectories\MC{"initial states from which $\pi$ does not satisfy $\phi$"?} fail under $\pi$
(``$\mathsf{f}$" stands for ``failure"). With the formalization above, we identify a feasible repair problem that takes account of trade-offs in initial states.
% is aware of initial states trade-offs.

\begin{definition}[\ourproblemfull{} Problem]
The \ourproblem{} problem is formalized as
\label{def:tagr_problem}
    \begin{equation}
    \label{eq:example_modified}
    \begin{split}
        &\text{maximize}_{\pi'}\, 
        \int_{\setfail{\pi}} \mathbbm{1}(\rob(s_0, \pi') \geq 0) ds_0\\
        &\text{subject to }\, \setsuccess{\pi} \subseteq \setsuccess{\pi'},
    \end{split}
\end{equation}
where $\mathbbm{1}(\cdot)$ is the indicator function\edit{, which returns 0 if the input proposition is false and 1 if true}.
\end{definition}
% \MC{To me, it's weird that the integral is over the unsafe states of $\pi$ instead of $\pi'$.}
% Here, $S_{\pi}^g \subset S_{init}$ is the subset of initial states that produces trajectories that are able to accomplish the task under controller $\pi$ (``g" stands for ``good") and $S_{\pi}^b = S_{init} \setminus S_{\pi}^g$ (``b stands for ``bad"). The goal of Equation \eqref{eq:example_modified} is to maximize the number of repaired bad initial states using an alternative controller $\pi'$, while protecting previously good initial states from being compromised, i.e., $|S_{\pi'}^b|$ is maximized under the constraint $S_{\pi}^g \subseteq S_{\pi'}^g$. Trivially, a controller that satisfies the constraint is the original controller $\pi$.

% In plain words, Equation \eqref{eq:example_modified} aims to find an alternative controller $\pi'$ to repair as large region of initial states as possible 
% \MC{I don't think that this is what Equation (7) is saying. It's trying to maximize the average robustness of all the unsafe states}. 
Equation \eqref{eq:example_modified} aims to find an alternative controller $\pi'$ to repair as many previously failed initial states in $\setfail{\pi}$ as possible\edit{, while protecting the} previously successful initial states $\setsuccess{\pi}$ from being compromised. Trivially, a controller that satisfies the constraint is the original controller $\pi$. Therefore, it is guaranteed that there exists a solution to this problem.

However, the objective function in Equation \eqref{eq:example_modified} contains a non-standard integral that is hard to compute. We hence approximate the problem by partitioning the initial state set into finitely many regions, i.e., $S_{init} = S_1 \cup \cdots \cup S_M$. There exists two types of initial state regions $S_i$: (i) successful: all states in the region are successful under a given controller $\pi$, i.e., $S_i \subseteq \setsuccess{\pi}$ and (ii) failed: not all states in the region are successful under $\pi$, i.e., $S_i \not\subseteq \setsuccess{\pi}$. The problem becomes
% \MC{I would call the regions 'successful' and 'unsuccessful'. The 'all-successful' and 'not all-successful' names are awkward. }
\begin{equation}
    \label{eq:example_discretized}
    \begin{split}
        &\text{maximize}_{\pi'} 
    \sum_{S_i \not\subseteq \setsuccess{\pi}} \mathbbm{1}(S_i \subseteq \setsuccess{\pi'}) \\
    &\text{subject to }\, S_i \subseteq \setsuccess{\pi} \implies S_i \subseteq \setsuccess{\pi'}.
    \end{split}
\end{equation}
The repair problem in Equation \eqref{eq:example_discretized} aims to repair as many failed regions as possible, while safeguarding the already successful ones. This discretized version of problem approaches Equation \eqref{eq:example_modified} as the partition becomes finer.

% However, solving this problem is still hard due to the uncountable size of $S_{init}$. We cannot obtain the exact partition $S_{\pi}^a \cup S_{\pi}^f$ nor design an algorithm to compute the solution due to infinitely many initial states. Therefore, we seek to discretize the state space.

\subsection{Feasible \ourproblem{} Problem Aided by Verification}
\label{subsec:main_problem}

To solve Equation \eqref{eq:example_discretized}, one more key issue remains: how do we know whether \textit{all states} in a region $S_i$ would lead to success? To provide guarantees for infinite (but bounded) regions, we hereby introduce the use of a sound but incomplete verifier, such as Verisig \cite{ivanov2019verisig,ivanov2021verisig}. This type of verification is conservative.
That is, if a region of initial states passes this verifier, it is guaranteed that all states in this region can successfully accomplish the task \edit{(i.e., the verifier is sound)}. However, the implication does not hold in the other direction \edit{(i.e., the verifier is incomplete)}. Formally, we have Definition \ref{def:verifier}.
% That is, if it verifies a region of initial states as all-successful, it is guaranteed that it is actually all-successful, but not vice versa. It may falsely verify all-successful regions as not all-successful.

\edit{
\begin{definition}[Sound but Incomplete Verifier]
    \label{def:verifier}
    Given dynamics $f$, a \emph{verifier} of property $\varphi$ is abstracted as a function $\veri: 2^{S} \times \Pi \mapsto \{0, 1\}$. 
    On a subset of initial states $S' \subseteq S$ for a given controller $\pi$, the verifier outputs whether all initial states in $S'$ produce trajectories that \edit{satisfy $\phi$}. A verifier is sound but incomplete iff
    \begin{equation}
        \label{eq:verifier}
        \begin{split}
            &\Big( \forall S' \in 2^{S} \cdot \veri(S', \pi) = 1 \implies S' \subseteq \setsuccess{\pi}\Big) ~\land\\ 
            &\Big( \exists S'' \in 2^{S} \cdot S'' \subseteq \setsuccess{\pi} \land %\implies 
            \veri(S'', \pi) = 0 \Big).
        \end{split}
    \end{equation}
\end{definition}
}
\edit{In Equation \eqref{eq:verifier}, the term on the first line means soundness, and the second line means incompleteness.}

% Equation \eqref{eq:verifier} is equivalent to soundness of the verifier. The other direction of this implication means completeness. Therefore, with a sound but incomplete verifier, the other direction does not necessarily hold.
% With Definition \ref{def:verifier}, we assume we have the following two tools.

% Based on the need to discretize state space. We can partition the initial state set $S_{init}$ into $M$ small regions $S_1 \cup S_2 \cup \cdots \cup S_M$. Assume we have the following tools.\\

% \begin{assumption}[Available Verifier]
% In addition to the two subroutines available in Assumption \ref{assum:available_subroutines}, we also have a sound but incomplete verifier $\veri: 2^\mathcal{S} \times \Pi \mapsto \{0, 1\}$ as in Definition \ref{def:verifier}. 
% \end{assumption}

% With the task checker and verifier available, we are able to estimate whether an initial state region $S_i \subseteq \setsuccess{\pi}$ by either (i) sampling multiple states in $S_i$ and calling the task checker, or (ii) running the verifier on the entire region. 

With a verifier available, we aim to address the following main problem, which first requires to partition the initial state space into regions \edit{$S_1, \dots, S_M$} 
% $S_{init} = S_1 \cup \dots \cup S_M$ \MC{This partition has already been defined}
. We know that a region $S_i$ is successful if it passes verification, i.e., $\veri(S_i, \pi_\theta) = 1$. We would like to protect such regions, preserving the correct behaviors. While protecting these regions, we would like to maximally repair failed regions. However, the \edit{incomplete property of the} verifier \edit{means that a repaired region may still be classified as unsafe by the verifier}. Consequently, we estimate a region $S_i$ is repaired if all initial states in a uniformly sampled finite subset $\hat{S}_i \subset S_i$ can accomplish the task, i.e., STL robustness $\geq 0$. \edit{The size of regions is a design choice. Since the verifier is conservative, the larger regions we have, the less likely we are able to capture the small subsets of initial states that are correct. However, larger regions would also lead to smaller numbers of regions, providing computational efficiency. Indeed, picking the sizes of regions exhibits an accuracy-efficiency trade-off. For simplicity, we assume the partition on regions is already given. Then} the main problem is formalized as follows.\\
% With the two tools available, we are able to check whether an initial state or a subset of initial states belongs to $S_{\pi}^a$ under some controller $\pi$. We therefore have the main problem.\\

\textbf{Main Problem} (Verification-aided \ourproblem{}). We would like to accomplish a given STL task $\varphi$ under dynamics $f: S \times A \mapsto S$. The current controller $\pi: S \mapsto A$ is unable to accomplish this task from all initial states in $S_{init} \subset S$. Therefore, we aim to find an alternative controller $\pi'$ by repairing a subset of initial states while safeguarding another. Specifically, we are given a partition $S_{init} = S_1 \cup \cdots \cup S_M$ of the initial state space. The regions to be repaired are
\begin{equation}
    \label{eq:to_be_repaired_regions}
    \mathcal{S}^{\mathsf{f}}_\pi := \{S_i : \exists s_0 \in \hat{S}_i, \rob(s_0, \pi) < 0\},
\end{equation}
where $\hat{S}_i \subset S_i$ is a uniformly and independently sampled finite subset from $S_i$, with size $K$. The regions to be protected are
% \MC{According to these two definitions, there may be regions in the middle that are neither safe nor unsafe (since the verifier is incomplete)}\EL{We could miss some regions that are ``actually" safe due to this incompleteness - we just need to repair the ones with failures sampled, while protecting the verified ones.}\MC{I think it would be good to mention this/. Maybe in a footnote if there's space.}
\begin{equation}
    \label{eq:to_be_protected_regions}
    \mathcal{S}^{\mathsf{s}}_\pi := \{S_i : \veri(S_i, \pi) = 1\}
\end{equation}
Assume we have a robustness computation subroutine $\rob: S \times \Pi \mapsto \mathbb{R}$ as in Definition \ref{def:robustness_computing} and a sound but incomplete verifier $\veri: 2^S \times \Pi \mapsto \{0, 1\}$ as in Definition \ref{def:verifier}. To find an alternative controller $\pi'$, we solve the following optimization problem\footnote{\edit{Due to incompleteness of the verifier, there may exist some regions that does not pass verification but no failures can be found. We do not know their safety for sure and thus they are neither protected nor repaired.}}.
% \MC{Is this the ultimate problem we want to state? It doesn't deal with the fact that the verifier can give conservative results. I think that introducing the three types of region (verified, unverified but empirically successful, and unverified and empirically unsuccessful) before stating this problem might be helpful.}
\begin{equation}
    \label{eq:main_problem}
    \begin{split}
        &\text{maximize}_{\pi'}\, \sum_{S_i \in \mathcal{S}^{\mathsf{f}}_\pi}\mathbbm{1}(\forall s_0 \in \hat{S}_i, \rob(s_0, \pi') \geq 0)\\
        &\text{subject to } S_i \in \mathcal{S}^{\mathsf{s}}_\pi \implies \veri(S_i, \pi') = 1.
    \end{split}
\end{equation}

That is, our goal is to design a repair algorithm that maximizes the number of repaired initial state regions and safeguard the already successful ones. The verifier provides a formal proof of constraint-satisfaction for a solution controller.

\edit{One remark is that this main problem seeks a single new control policy $\pi'$. There exists an alternative problem setting, which aims to obtain one policy per region $S_i$ and switch between these policies at runtime. Although one model per region could provide high performance, retaining such a large number of models would require an unbounded large memory overhead, causing large latency, and energy during computation \cite{catthoor2013custom}. In contrast, researchers in multi-task and continual learning have already shown satisfactory performance in using one shared model for multiple sub-tasks \cite{ruvolo2013active}. Therefore, our research targets the solution of a shared controller representation. It would be interesting for future works to explore the alternatives.}

\section{\ourmethodfull{} Algorithm}
\label{sec:method}

We design an algorithm for the main problem. First, in Section \ref{subsec:energy_function}, we design an optimization objective function (energy function) to be maximized. This energy function aims to improve the STL robustness of one set of initial states while protecting the positive robustness of another set of initial states. In Section \ref{subsec:safe_sim_annealing}, we design a safeguarded version of simulated annealing to optimize the energy function. Finally, in Section \ref{subsec:ccgr} we propose the overall \ourmethod{} algorithm that uses the safeguarded simulated annealing as a subroutine to incrementally repair failed state regions.

\subsection{Energy Function Design}
\label{subsec:energy_function}

% We denote the controller parameters as $\theta \in \Theta$\MC{Should this be defined when we define the controller $\pi$?}.\edit{In our case}, $\Theta$ is the set of all parameters of a neural network. A controller $\pi$ parameterized by $\theta$ is denoted as $\pi_\theta$. 
We will use a simulated annealing-based approach to optimize the \edit{parameters $\theta$ for a controller $\pi_\theta$}, due to its ability to escape local optima. 

To apply simulated annealing, our first step is to design the energy function.
\edit{To repair a failed region of initial states $S_i \in \mathcal{S}^{\mathsf{f}}_\pi$ (as defined in the main problem), we aim to improve the STL robustness of its generated trajectories. This results in the following energy function}
\edit{
\begin{equation}
    \label{eq:energy_1}
    e(\theta) := \frac{1}{|S_i|}\int_{S_i} \rob(s_0, \pi_\theta) ds_0.
\end{equation}
}\edit{Here, $|\cdot|$ denotes a volumetric measure on a continuous set, $|S| = \int_{S} ds_0$, i.e., the volume if the set $S$ is Euclidean ($S \subset \mathbb{R}^n$).} That is, we want to maximize the \edit{average} STL robustness of trajectories from all initial states $s_0 \in S_i$.

However, this energy function does not protect any currently successful initial states. Formally, we want to repair a region $S_i\in \mathcal{S}^{\mathsf{f}}_\pi$ \edit{while safeguarding $S^{\mathsf{s}} := \bigcup_{S \in \mathcal{S}^{\mathsf{s}}_\pi} S$}, as \edit{stated in} the main problem.
% \MC{What are $S_1$ and $S_2$ here? We should link them to the successfull/unsuccessful dichotomy we defined earlier.} 
\edit{To do this, we employ a} log-barrier function, as it is a smooth approximation of \edit{the constraint} $\forall s_0 \in S^{\mathsf{s}}$, $\rob(s_0, \pi_\theta) \geq 0$ and is easy to compute \cite{polyak1992modified,den1992classical,hauser2006barrier}. We have
% \MC{Why is (12) different from the first line of (13). They should be the exact same}
\begin{equation}
    \label{eq:energy_2}
    \begin{split}
        e(\theta) &:= \frac{1}{|S_i|}\int_{S_i} \rob(s_0, \pi_\theta) ds_0\\
        &+ \frac{\lambda}{|S^{\mathsf{s}}|}\int_{S^{\mathsf{s}}}\log(\rob(s_0, \pi_\theta)) ds_0
    \end{split}
\end{equation}
The hyperparameter $\lambda > 0$ is a balance factor. Larger values of $\lambda$ favor protecting the successful states, while smaller values favor fixing the unsuccessful states. The energy function in Equation \eqref{eq:energy_2} aims to improve the robustness of all initial states in $S_i$, while safeguarding the robustness of all initial states in $S^{\mathsf{s}}$ with log barrier. \edit{Note that if the STL robustness is negative then the logarithm will be undefined. To avoid this, we set up a lower bound, e.g. $-1000$, that the log-robustness will take if the STL robustness is negative or if log robustness is lower than the bound.}

% One computational issue is that evaluating Equation \eqref{eq:energy_2} requires non-standard integration, which is impractical.
One computational issue is that Equation \eqref{eq:energy_2} has no convenient closed-form solution, making it hard to evaluate directly.
In response, we use Monte Carlo integration \cite{robert1999monte} to approximate the two integrals. 
So, we uniformly sample states from each region given in the main problem, which we denote as $\hat{S}_i \subset S_i$ and $\hat{S}^{\mathsf{s}} \subset S^{\mathsf{s}}$, respectively. The cardinality of $\hat{S}_i$ and $\hat{S}^{\mathsf{s}}$ are $K$ and $LK$ \edit{($K$ samples per region and $L=|\mathcal{S}^{\mathsf{s}}_\pi|$ regions to protect)}. \edit{So the energy function now becomes}
% we uniformly and independently sample $K_1$ initial states in $S_1$ and $K_2$ initial states in $S_2$. Denote the two finitely sampled sets of initial states as $\hat{S}_1 \subset S_1$ and $\hat{S}_2 \subset S_2$, respectively. We have
\begin{equation}
    \label{eq:energy_3}
    \begin{split}
        \hat{e}(\theta) &:= \frac{1}{K}\sum_{s_0 \in \hat{S}_i} \rob(s_0, \pi_\theta)\\
        & + \frac{\lambda}{LK}\sum_{s_0 \in \hat{S}^{\mathsf{s}}} \log(\rob(s_0, \pi_\theta)).
    \end{split}
\end{equation}

% We would like to quantify the gap between $\hat{e}(\theta)$ and $e(\theta)$, in order to show that they can be arbitrarily close. The following proposition holds.
The following proposition shows that $\hat{e}(\theta)$ and $e(\theta)$ can be arbitrarily close.

\begin{proposition}[Error in Monte Carlo Integrated Energy Function]
\label{prop:bounded_error}
The energy functions $\hat{e}(\theta)$ in Equation \eqref{eq:energy_3} and $e(\theta)$ in Equation \eqref{eq:energy_2} satisfy
\begin{equation}
    \begin{split}
        &\mathbb{E}[\hat{e}(\theta)] = e(\theta) \text{ and}\\
        &\text{Var}[\hat{e}(\theta)] = \left(\frac{1}{K}\right)^2 \text{Var}_{\hat{S}_i}[\rob] + \left(\frac{\lambda}{LK}\right)^2 \text{Var}_{\hat{S}^{\mathsf{s}}}[\log\circ\rob].
    \end{split}
\end{equation}
Here, $\text{Var}_{\hat{X}}[g]$ denotes the variance of a function $g$ on a finite set of inputs $\hat{X}$.
\end{proposition}

\edit{Our proof for Proposition~\ref{prop:bounded_error} is provided in the Appendix.} Proposition~\ref{prop:bounded_error} states that the Monte Carlo integrated energy function $\hat{e}(\theta)$ has an expected value of the energy function $\hat{e}$. Moreover, as we sample more initial states (larger $K$), the variance of $\hat{e}(\theta)$ decreases. Therefore, by increasing $K$, the approximation can be arbitrarily close to the original energy function. Specifically, we are able to compute confidence intervals based on the size of sampled state sets \cite{preacher2012advantages}.

\subsection{Safeguarded Simulated Annealing}
\label{subsec:safe_sim_annealing}

% With a computable energy function defined in Equation \eqref{eq:energy_3}, we here propose a safeguarded version of simulated annealing.

The energy function in Equation \eqref{eq:energy_3} is a non-standard function with a potentially large number of local optima. Therefore, we utilize simulated annealing \edit{to maximize it}, which has an advantage of escaping local optima as mentioned in Section \ref{subsec:sim_annealing}. However, due to its stochastic nature, simulated annealing may compromise the protected initial states, lowering their trajectories' STL robustness. We tolerate such compromise unless the robustness drops below 0, which means the task is failed. Therefore, we propose the following safeguarded version of simulated annealing in Algorithm \ref{alg:safe_sa}, which has an additional check on the STL robustness of the protected initial states.

\begin{algorithm}
\caption{Safeguarded Simulated Annealing}
\label{alg:safe_sa}
\hspace*{\algorithmicindent} \textbf{Input}: Finite set of states to be repaired $\hat{S}_i$, finite set of states to be protected $\hat{S}^{\mathsf{s}}$, controller parameters $\theta$, energy balance factor $\lambda > 0$, perturbation standard deviation $\sigma > 0$, initial temperature $\tau > 0$, cooling factor $\alpha \in (0, 1)$, max iteration $maxIter$\\
\hspace*{\algorithmicindent} \textbf{Output}: Intermediate repaired controller parameters $\theta$
\begin{algorithmic}[1]
\State $\hat{e}, \rho_{\hat{S}^{\mathsf{s}}} \gets \operatorname{evaluateEnergy}(\hat{S}_i, \hat{S}^{\mathsf{s}}, \theta, \lambda)$
\For{$iter = 1, \dots, maxIter$}
    \State $\theta' \gets \theta + \mathcal{N}(0, \sigma^2)$
    \State $\hat{e}', \rho_{\hat{S}^{\mathsf{s}}} \gets \operatorname{evaluateEnergy}(\hat{S}_i, \hat{S}^{\mathsf{s}}, \theta', \lambda)$
    \State $\edit{\Delta_{\hat{e}}} \gets \hat{e}' - \hat{e}$
    \State Draw $x \sim Bernoulli (\exp(-\Delta \hat{e} / \tau))$
    \If{$\edit{\Delta_{\hat{e}}} \geq 0$ or $x = 1$}
        \If{$\rho_{\hat{S}^{\mathsf{s}}} \geq 0$}
            \State $\theta \gets \theta'$, $\hat{e} \gets \hat{e}'$
        \EndIf
    \EndIf
    \State $\tau \gets \tau \times \alpha$
\EndFor
\end{algorithmic}
\end{algorithm}

Algorithm \ref{alg:safe_sa} presents a safeguarded version of simulated annealing. Specifically, on a given $S_i$ to be repaired, it aims to improve robustness on the finite sampled initial set $\hat{S}_i$ while protecting the robustness of $\hat{S}^{\mathsf{s}}$. The key subroutine is $\operatorname{evaluateEnergy}$ at lines 1 and 4\edit{, which takes as input $\hat{S}_i$, $\hat{S}^{\mathsf{s}}$, current controller parameters $\theta$, and balance factor $\lambda$} and outputs energy $\hat{e}$ as in Equation \eqref{eq:energy_3} and the minimum robustness $\rho_{\hat{S}^{\mathsf{s}}}$ of protected initial states, i.e.,
\begin{equation}
    \label{eq:min_rob}
    \rho_{\hat{S}^{\mathsf{s}}}(\theta) := \min_{s_0 \in \hat{S}^{\mathsf{s}}} \rob(s_0, \pi_\theta).
\end{equation}
% We can see this can be computed along with the energy in Equation \eqref{eq:energy_3}, which requires calling the robustness computation $\rob$ on all $s_0 \in \hat{S}_2$.\MC{I don't think we need this sentence.}

From lines 2 to 13 we have the main loop of simulated annealing. The parameters are perturbed with Gaussian noise at line 3, and energy is evaluated again at line 4. Line 7 provides the standard criterion check of simulated annealing as explained in Section \ref{subsec:sim_annealing}, i.e., Metropolis-Hastings criterion. However, this criterion check is insufficient to safeguard all initial states in $\hat{S}^{\mathsf{s}}$, since a new parameter $\theta'$ may break them. Hence, we add an additional safeguard on the minimum robustness at line 8, to ensure that all $s_0 \in \hat{S}^{\mathsf{s}}$ still produce trajectories that accomplish the task. \edit{Finally}, the temperature cools down at line 13 to gradually discourage exploration.

\subsection{The \ourmethod{} Algorithm}
\label{subsec:ccgr}

With the safeguarded simulated annealing defined in Algorithm \ref{alg:safe_sa}, we use it as a subroutine in our main \ourmethod{} algorithm, detailed in Algorithm \ref{alg:ccgr}.

\begin{algorithm}
\caption{\ourmethodfull{}}
\label{alg:ccgr}
\hspace*{\algorithmicindent} \textbf{Input}: Partitioned continuous initial state set $S_{init} = S_1 \cup \dots \cup S_M$, controller parameters $\theta$, sample size $K$, energy balance factor $\lambda > 0$, perturbation std $\sigma > 0$, initial temperature $\tau > 0$, cooling factor $\alpha \in (0, 1)$, max iteration $maxIter$\\
\hspace*{\algorithmicindent} \textbf{Output}: Repaired controller parameters $\theta$ and final verification outputs $v_1', \dots, v_M'$
\begin{algorithmic}[1]
\For{each $S_i$}
    \State $v_i \gets \veri(S_i, \pi_\theta)$
\EndFor
\For{each $S_i$}
    \State Uniformly sample $K$ states in $S_i$ to form finite set $\hat{S}_i$
    \For{each $s_0^{j} \in \hat{S}_i$}
        \State $\rho_i^j \gets \rob(s_0^{j}, \pi_\theta)$
    \EndFor
\EndFor
\State $\mathcal{S}^{\mathsf{s}} \gets \{S_i : v_i = 1\}$ \Comment{regions to be protected}
\State $\hat{S}^{\mathsf{s}} \gets \bigcup_{S_i \in \mathcal{S}^{\mathsf{s}}} \hat{S}_i$
\State $\mathcal{S}^{\mathsf{f}} \gets \{S_i : \exists j, \rho_i^j < 0\}$ \Comment{regions to be repaired}
\State Sort $\mathcal{S}^{\mathsf{f}}$ in decreasing order of $\sum_j \rho_i^j$ of each $S_i$
\While{$\mathcal{S}^{\mathsf{f}}$ is not empty}
    \State $S_i \gets$ first region in $\mathcal{S}^{\mathsf{f}}$ in sorted order
    \State $\hat{S}^{\mathsf{f}} \gets \{s_0^j \in \hat{S}_i : \rho_i^j < 0\}$ 
    \State $\theta' \gets \operatorname{safeSimAnnealing}(\hat{S}^{\mathsf{f}}, \hat{S}^{\mathsf{s}}, \theta, \lambda, \sigma, \tau, \alpha, maxIter)$
    % \State $\mathcal{S}^{\mathsf{f}} \gets \mathcal{S}^{\mathsf{f}} \setminus \{S_i\}$
    \If{$\theta' \neq \theta$} \Comment{new controller identified}
        \For{each $S_k \in \mathcal{S}^{\mathsf{f}}$} \Comment{re-evaluate robustness}
            \For{each $s_0^j \in S_k$}
                \State $\rho^j_k \gets \rob(s_0^j, \pi_{\theta'})$
            \EndFor
        \EndFor
        \State Sort $\mathcal{S}^{\mathsf{f}}$ in decreasing order of $\sum_j \rho_k^j$ of each $S_k$
        \State $\mathcal{S}' \gets \{S_k \in \mathcal{S}^{\mathsf{f}} : \min_j \rho_k^j \geq 0\}$ \Comment{newly repaired regions}
        \State $\mathcal{S}^{\mathsf{f}} \gets \mathcal{S}^{\mathsf{f}} \setminus \mathcal{S}'$, $\mathcal{S}^{\mathsf{s}} \gets \mathcal{S}^{\mathsf{s}} \cup \mathcal{S}'$
        \State $\hat{S}^{\mathsf{s}} \gets \bigcup_{S_k \in \mathcal{S}^{\mathsf{s}}} \hat{S}_k$, $\theta \gets \theta'$
    \EndIf
\EndWhile
\For{each $S_i$}
    \State $v_i' \gets \veri(S_i, \pi_\theta)$
\EndFor
\end{algorithmic}
\end{algorithm}

Algorithm \ref{alg:ccgr} can be divided into three parts: preparation (lines 1-13), main repair loop (lines \edit{14-29}) and final verification (lines \edit{30-32}). In the preparation phase each initial state region $S_i$ is verified under current controller $\pi_\theta$ by calling the sound but incomplete verifier $\veri$. This step can be parallelized on individual regions. At lines 4-9 we uniformly sample $K$ initial states from each region $S_i$ to form a finite set, in order to estimate the energy function via Monte Carlo integrals. The robustness of each sampled initial state is evaluated at line 7. This sampling and robustness computation can \edit{also be parallelized}.

\edit{The regions and states to be protected are defined in lines 11 and 12, respectively, while the regions to be repaired are defined in line 12. As a final preparation step, we sort the to-be-repaired regions by decreasing order of average sampled robustness, since we expect regions with high robustness to be easier to repair. }

%The regions to be protected are all regions that pass the verification, and their sampled states will be protected during safeguarded simulated annealing (line 10-11). The regions to be repaired are the regions with failed sampled states, i.e., robustness $\rho_{i}^j < 0$ (line 12). At line 13, we sort the to-be-repaired regions by decreasing order of average sampled robustness. That is, we want to repair regions from high robustness to low, because an already high robustness implies it is relatively easy to be repaired.

In the main repair loop from line 14 to \edit{29}, \edit{we first select the next failed region to repair (line 14) and identify which of its sampled states in $\hat{S}_i$ fail to accomplish the task. } These states then get repaired by the safeguarded simulated annealing (Algorithm \ref{alg:safe_sa} called at line 17). \edit{After simulated annealing, we check whether a new controller is obtained.} 
% After the simulated annealing procedure, the current region being repaired is removed from the to-be-repaired set (line 18). Then, we check whether a new controller is obtained.
This information can be passed down from Algorithm \ref{alg:safe_sa}. If we have a new controller, robustness of the sampled states in the remaining to-be-repaired regions are evaluated again (lines \edit{19-23}), and the regions are sorted based on the evaluation (line \edit{24}). Repairing a region $S_i$ may also repair other regions, so we identify all the repaired regions as $\mathcal{S}'$ (line \edit{25}). These regions are taken away from to-be-repaired regions and added to to-be-protected ones (line \edit{26}), so that we safeguard both the previously verified regions and the newly repaired. 

Finally, after the repair is done, we rerun the verifier at lines \edit{30-32} to obtain the final verification results. The verifier is not called during the repair procedure because it is generally expensive compared to computing robustness on individual sampled states. 

We identify two potential challenges in execution of Algorithm \ref{alg:ccgr}. First, there may exist a protected region that does not pass verification at the end. One thing we can do is to increase the number of protected sampled states (larger $K$) in that region, or to adjust the balance factor $\lambda$. However, we did not encounter this case during case studies. Second, the main repair loop can take a long time. In practice, this loop can be terminated early when the repairing speed is slow, i.e., very few to no additional regions are repaired and removed at line 26. This observation implies that $\theta$ is converged.

\section{Case Studies}
\label{sec:case_studies}

\subsection{Experiment Setup}
\label{subsec:setup}

% We perform two case studies as follows. 
We perform two case studies. In both, the hyperparameters in Algorithm \ref{alg:ccgr} are $K=100$, $\lambda = 1$, $\sigma = 0.01$, $\tau = 1$, $\alpha = 0.95$, $maxIter = 100$. 

We compare \ourmethod{} with two baselines: \footnote{We reached out to the authors of another relevant technique~\cite{zhou2020runtime} to use it as a baseline but did not receive the source code in time for this submission.}
\begin{enumerate}
    \item \textbf{Gradient ascent:} When maximizing the energy function in Equation \eqref{eq:energy_3}, this baseline performs gradient ascent instead of simulated annealing. To compute the gradient $\partial{\hat{e}}/\partial{\theta}$, we soften the STL robustness function to make the energy function differentiable, as per the existing literature~\cite{haghighi2019control}. Then the controller parameter $\theta$ is updated as $\theta' \gets \theta + \eta \times (\partial{\hat{e}}/\partial{\theta})$, with step size $\eta \in \{0.01, 0.001, 0.0001\}$. A new controller is accepted only if it does not lower the protected states' robustness below 0.
    \item \textbf{Non-safeguarded simulated annealing:} This baseline maximizes an energy function without log barrier, i.e., the Monte Carlo integral approximation of Equation \eqref{eq:energy_1}. No initial states are protected.
\end{enumerate}

The following concepts help easily parse our results.
\begin{enumerate}
    \item $\mathcal{S}^{\mathsf{s}}_\pi$: the set of regions that pass verification under a controller $\pi$ (same as Equation \eqref{eq:to_be_protected_regions}).
    \item $\tilde{\mathcal{S}}^{\mathsf{s}}_\pi$: the set of regions that do not pass verification under a controller $\pi$, but for which no failure is identified by sampling $K=100$ initial states per region.
    \item $\mathcal{S}^{\mathsf{f}}_\pi$: the set of regions with failure sampled under a controller $\pi$ (same as Equation \eqref{eq:to_be_repaired_regions}).
\end{enumerate}

We track three metrics according to the main problem. First, we check how many verified regions are lost, i.e., the number of regions in $\mathcal{S}^{\mathsf{s}}_\pi$ that no longer pass verification. Second, we check how many broken regions are repaired, i.e., the number of regions in $\mathcal{S}^{\mathsf{f}}_\pi$ that no longer contain any failed sampled states. Third, we evaluate the STL robustness of the sampled states. We use Verisig \cite{ivanov2019verisig,ivanov2021verisig} as our verifier.

The experiments are run on Intel(R) Xeon(R) Gold 6148 CPU @ 2.40Hz. All parallel computations are distributed to CPUs of this type. The OS is Ubuntu 20.04.6 LTS, with kernel Linux 5.4.0-159-generic and architecture x86-64. \edit{We provide our code in a repeatability package, detailed in the Appendix.}

\subsection{Case Study 1: Unmanned Underwater Vehicle}
\label{subsec:uuv}

\begin{table*}[]
\normalsize
\centering
\caption{UUV Repair Results ($\pi' = \pi$ Before Repair)}
\begin{tabular}{|c|c|c|c|c|c|c|}
\hline
 & $|\mathcal{S}^{\mathsf{s}}_{\pi'}|$ : $|\tilde{\mathcal{S}}^{\mathsf{s}}_{\pi'}|$ : $|\mathcal{S}^{\mathsf{f}}_{\pi'}|$ & \begin{tabular}[c]{@{}c@{}}\# of regions\\in $\mathcal{S}^{\mathsf{s}}_{\pi}$ broken\end{tabular} & \begin{tabular}[c]{@{}c@{}}\# of regions\\in $\mathcal{S}^{\mathsf{f}}_{\pi}$ repaired\end{tabular} & \begin{tabular}[c]{@{}c@{}}Min rob\\per region\\in $\mathcal{S}^{\mathsf{f}}_{\pi'}$\end{tabular} & \begin{tabular}[c]{@{}c@{}}Min rob\\per region\\in $\mathcal{S}^{\mathsf{s}}_{\pi'} \cup \tilde{\mathcal{S}}^{\mathsf{s}}_{\pi'}$\end{tabular} & \begin{tabular}[c]{@{}c@{}}Min rob\\per region\\overall\end{tabular} \\ \hline
 Before repair & $141:963:896$ & N/A & N/A & $-0.24 \pm 0.06$ & $2.79 \pm 1.89$ & $1.49 \pm 2.05$ \\ \hline
Gradient ascent &$141:963:896$& 0 (0\%) & 0 (0\%) & $-0.24 \pm 0.06$ & $2.79 \pm 1.89$ & $1.49 \pm 2.05$ \\ \hline
\begin{tabular}[c]{@{}c@{}}Non-safeguarded\\ sim annealing\end{tabular} &$132:1702:166$& 9 (6.4\%) & 730 (81.4\%) & $-0.1 \pm 0.03$ & $4.73 \pm 2.79$ & $4.27 \pm 3.01$ \\ \hline
\ourmethod{} (ours) &$173:1661:166$& 0 (0\%) & 730 (81.4\%) & $-0.1 \pm 0.04$ & $4.92 \pm 2.8$ & $4.51 \pm 3.02$ \\ \hline
\end{tabular}
\label{tab:uuv_result}
\end{table*}

The Unmanned Underwater Vehicle (UUV) control problem is based on a challenge problem from the DARPA Assured Autonomy program~\cite{ruchkin2022confidence,ivanov2021verisig}. The UUV has a four-dimensional state space $(x_t, y_t, h_t, v_t)$, where $(x_t, y_t)$ are the two-dimensional coordinates, $h_t$ is the heading angle and $v_t$ is the velocity, as illustrated in Figure \ref{fig:uuv}. The $x$-coordinate always starts at $0$, and the velocity is fixed at $0.4855$ m/s.

\begin{figure}[]
\normalsize
    \centering
    \includegraphics[width=0.35\textwidth]{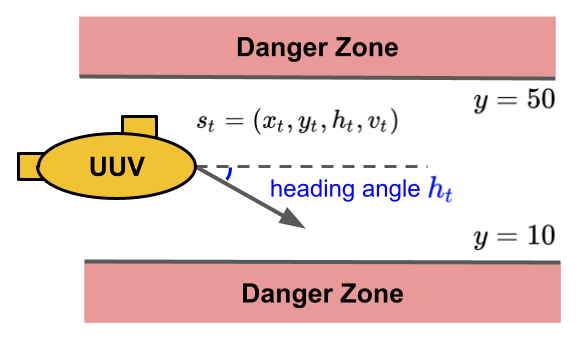}
    \caption{Unmanned Underwater Vehicle}
    \label{fig:uuv}
\end{figure}

The UUV must travel within a range of distances from a pipe that it is scanning. The upper and lower distance bounds are at $y = 10$ and $y = 50$, respectively. The controller's goal is to keep the UUV within this safe range for the next 30 seconds. We formalize this task in STL as
\begin{equation}
    \label{eq:uuv_goal}
    \varphi = G_{t \in [0, 30]} ((y_t > 10) \land (y_t < 50)).
\end{equation}
Every second, the UUV gets two measurements: the heading angle to the pipe and the distance to the lower edge of pipe. We assume the measurement noise is negligible. The UUV then computes a one-dimensional turning angle action $\Delta h_t$ based on the two inputs via a neural network controller, which has 2 hidden layers with 32 neurons each. Each layer is tanh-activated~\cite{ivanov2021verisig,ruchkin2022confidence}.

The initial state space of the UUV is $S_{init} = \{(y, h) \mid y \in [12, 22], h \in [10, 30]\}$. To repair the controller, we partition $S_{init}$ into rectangular regions, with step size $0.1$m for $y$ and $1.0$ degree for $h$. As a result, we obtained $2000$ regions in total.

%We aim to repair a feed-forward neural network controller $\pi$ \cite{ivanov2021verisig,ruchkin2022confidence}, which has 2 hidden layer with 32 neurons each. Each layer is tanh-activated. We consider the two-dimensional initial state space $S_{init} = \{(y, h) : y \in [12, 22], h \in [10, 30]\}$. To run \ourmethod{}, this initial state space is partitioned into multiple rectangular regions, with step size = 0.1 for $y$ and 1.0 for $h$. Therefore, there are $2000$ regions in total.

\begin{table}[]
\caption{Means and standard deviations of\\ \ourmethod{} computation times in the UUV case study}
\normalsize
\centering
\begin{tabular}{|@{}p{1.3cm}@{}|c|c|c|}
\hline
           & \begin{tabular}[c]{@{}c@{}}Verification \\ (per region)\end{tabular} & \begin{tabular}[c]{@{}c@{}}Sim annealing\\ (per iter)\end{tabular} & \begin{tabular}[c]{@{}c@{}}Rob. check\\ (per iter)\end{tabular} \\ \hline
Time (s) & $233.29 \pm 262.69$                                                  & $180.63 \pm 13.1$                                             & $50.2 \pm 45.1$                                             \\ \hline
\end{tabular}
\label{tab:uuv_time}
\end{table}

% Before repair, there are 141 initial state regions that pass verification, 963 regions that fail verification but have no failures in 100 uniformly sampled initial states, and 896 regions that fail verification and contain unsafe initial states within 100 uniform samples. Table \ref{tab:uuv_result} shows the results of repair. 

With an alternative $\pi'$ identified for each method, Table \ref{tab:uuv_result} shows the repair results.
First, the gradient ascent method fails to identify new controller parameters without lowering at least one protected initial state's robustness below 0 under all step sizes $\eta$. In other words, it cannot escape its local optimum without compromising existing correct states. This shows the STL robustness's sensitivity to controller parameters. 
%For both non-safeguarded simulated annealing and \ourmethod{}, we are able to repair 730 of the 963 regions, such that no failure can be identified from sampling.
For both non-safeguarded simulated annealing and \ourmethod{}, we are able to repair $81.4\%$ of regions with failure. 
However, due to a lack of protection, this baseline is unable to preserve all the verified regions, with 9 verified regions becoming broken. On the other hand, \ourmethod{} not only protects but also increments the verified regions from 141 to 173, as illustrated in the left column of Figure \ref{fig:repair_results}. Moreover, \ourmethod{} also obtains higher STL robustness due to its protection. Here, we evaluate the minimum sampled robustness per region, with \ourmethod{} improving the overall score from 1.49 to 4.51 on average --- approximately 3 times.

\begin{figure}
    \centering
    \includegraphics[width=0.5\textwidth]{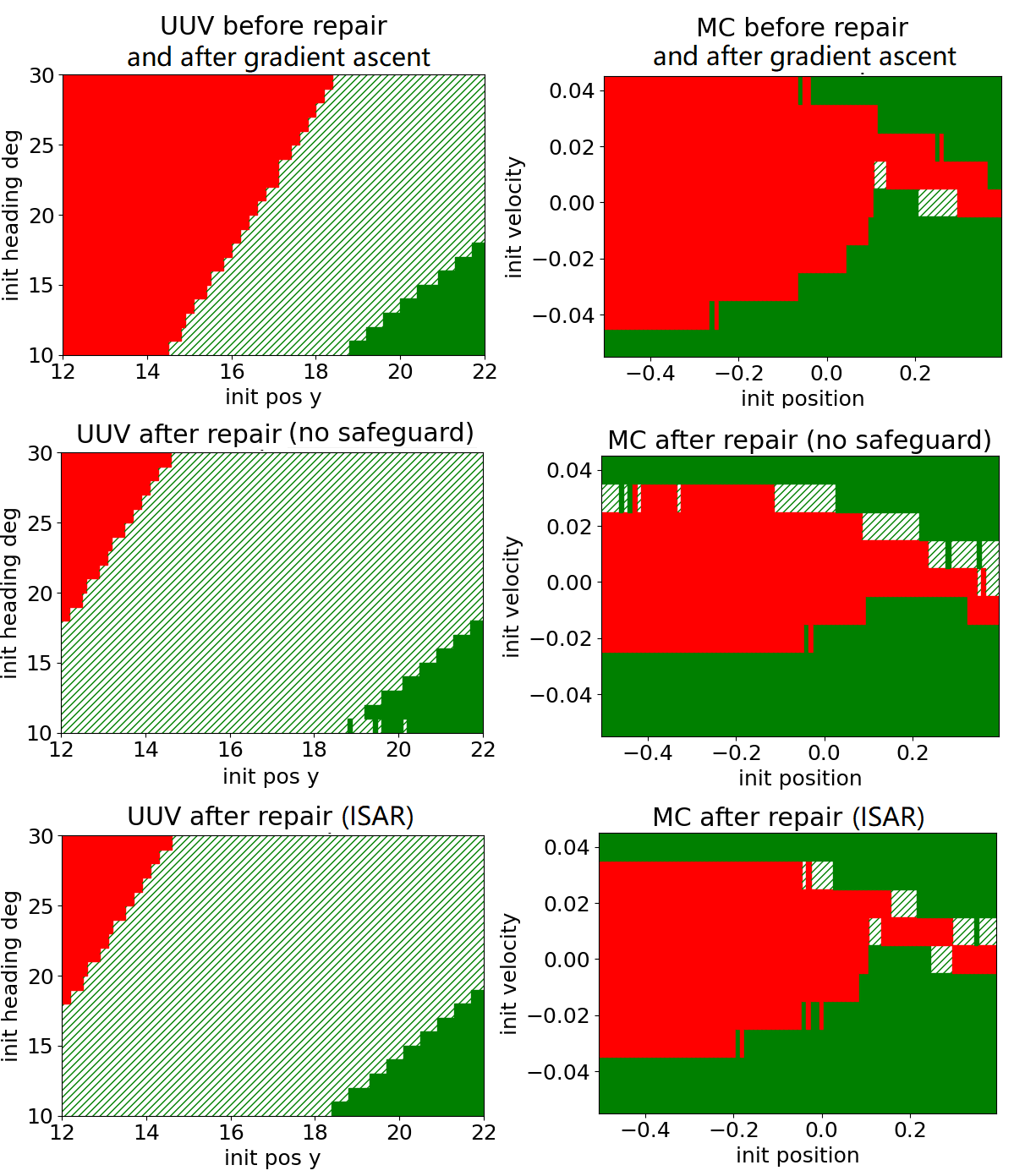}
    \caption{Repair results on initial state spaces of UUV (left column) and MC (right column). Solid green: regions in $\mathcal{S}_{\pi'}^{\mathsf{s}}$; green hatches: regions in $\mathcal{S}_{\pi'}^{\mathsf{s}'}$; red: regions in $\mathcal{S}_{\pi'}^{\mathsf{f}}$.}
    \label{fig:repair_results}
\end{figure}

\begin{table*}[]
\normalsize
\centering
\caption{Mountain Car Repair Results ($\pi' = \pi$ Before Repair)}
\begin{tabular}{|c|c|c|c|c|c|c|}
\hline
 & $|\mathcal{S}^{\mathsf{s}}_{\pi'}|$ : $|\tilde{\mathcal{S}}^{\mathsf{s}}_{\pi'}|$ : $|\mathcal{S}^{\mathsf{f}}_{\pi'}|$ & \begin{tabular}[c]{@{}c@{}}\# of regions\\in $\mathcal{S}^{\mathsf{s}}_{\pi}$ broken\end{tabular} & \begin{tabular}[c]{@{}c@{}}\# of regions\\in $\mathcal{S}^{\mathsf{f}}_{\pi}$ repaired\end{tabular} & \begin{tabular}[c]{@{}c@{}}Min rob\\per region\\in $\mathcal{S}^{\mathsf{f}}_{\pi'}$\end{tabular} & \begin{tabular}[c]{@{}c@{}}Min rob\\per region\\in $\mathcal{S}^{\mathsf{s}}_{\pi'} \cup \tilde{\mathcal{S}}^{\mathsf{s}}_{\pi'}$\end{tabular} & \begin{tabular}[c]{@{}c@{}}Min rob\\per region\\overall\end{tabular} \\ \hline
Before repair &$366 : 11 : 523$ & N/A & N/A & $ -0.37 \pm 0.22 $ & $ 0.14 \pm 0.02$ & $ -0.12 \pm 0.3$ \\ \hline
Gradient ascent&$366 : 11 : 523$ & 0 (0\%) & 0 (0\%) & $ -0.37 \pm 0.22 $ & $ 0.14 \pm 0.02$ & $ -0.12 \pm 0.3$ \\ \hline
\begin{tabular}[c]{@{}c@{}}Non-safeguarded\\ sim annealing\end{tabular} & $485 : 52 : 363$& 29 (7.9\%) & 189 (36.1\%) & $-0.39 \pm 0.19$ & $0.14 \pm 0.02$ & $ -0.05 \pm 0.29$ \\ \hline
\ourmethod{} (ours) &$469:21:410$ & 0 (0\%) & 123 (23.5\%) & $ -0.39 \pm 0.17 $ & $ 0.15 \pm 0.01$ & $ -0.03 \pm 0.27$ \\ \hline
\end{tabular}
\label{tab:mc_result}
\end{table*}

\begin{table}[!htbp]
\normalsize
\centering
\caption{Means and standard deviations of\\\ourmethod{} computation times in the MC case study}
\begin{tabular}{|@{}p{1.3cm}@{}|c|c|c|}
\hline
           & \begin{tabular}[c]{@{}c@{}}Verification \\ (per region)\end{tabular} & \begin{tabular}[c]{@{}c@{}}Sim annealing\\ (per iter)\end{tabular} & \begin{tabular}[c]{@{}c@{}}Rob. check\\ (per iter)\end{tabular} \\ \hline
Time (s) & $1959.58 \pm 557.5$ & $155.27 \pm 10.5$& $24.52 \pm 23.6$\\ \hline
\end{tabular}
\label{tab:mc_time}
\end{table}

In terms of computation time, repairing the UUV controller takes 7 main repair loops in the \ourmethod{} algorithm (lines 14-30 in Algorithm~\ref{alg:ccgr}). The major temporal overheads are recorded in Table~\ref{tab:uuv_time}. Here, verification is distributed to multiple CPU cores, with one thread per region. Calling the safeguarded simulated annealing subroutine is another major overhead, with the computational overhead relatively constant at each iteration. A third overhead is checking the robustness scores on all sampled states in all regions (line 20-24). With the number of states to be checked decreasing per iteration, this overhead also decreases.

\subsection{Case Study 2: Mountain Car}
\label{subsec:mc}

Mountain Car (MC) is a control problem in OpenAI Gym \cite{brockman2016openai}. Here, we control a car with a two-dimensional state space $(x_t, v_t)$, where $x_t$ is the one-dimensional coordinate (in left-right direction) and $v_t$ is the velocity. The goal is to drive the car from the bottom of a valley to the top of a mountain to its right ($x \geq 0.45$) within 110 seconds, formalized in STL as
\begin{equation}
    \varphi = F_{t \in [0, 110]}(x_t \geq 0.45).
\end{equation}
% This target-reaching problem is challenging because, based on the dynamics, the car can only reach the right mountain top if it first backs up to another mountain on its left to accumulate momentum.

For MC, we repair a feed-forward neural network controller $\pi$ \cite{ruchkin2022confidence}, with 2 hidden layers of 16 neurons each. The hidden layers are sigmoid-activated and the output layer is tanh-activated. The initial state space is $S_{init} = \{(x, v) : x \in [-0.505, 0.395], v \in [-0.055, 0.045]\}$, with partition step sizes of $0.1$ and $0.01$, respectively. There are 900 regions in total.

% Before repair, the three types of regions (verified, not verified but no sampled failures, not verified and containing sampled failures) are 366 : 11 : 523. 
Like with the UUV, we list the results in Table \ref{tab:mc_result}. Again, gradient ascent is unable to escape the local optimum and does not find an alternative controller. 
% The non-safeguarded baseline ends up with 485 : 52 : 363 of the three region types, and \ourmethod{} 469 : 21 : 410. 
Although non-safeguarded simulated annealing repairs more failed regions, the cost is breaking 29 verified ones, whereas \ourmethod{} does not break any. The regions are illustrated in the right column of Figure \ref{fig:repair_results}. The maximal STL robustness achievable in this case study is 0.15, since the largest $x$-coordinate is 0.6 and the robustness score is the maximal value of $x_t - 0.45$. Overall, \ourmethod{} is able to achieve higher STL robustness on average. The computation times are shown in Table \ref{tab:mc_time}. The verification time is longer than UUV due to the longer time horizon of 110 seconds.

\section{Discussion and Conclusion}
\label{sec:discussion}

% \textbf{Arbitrary State Selection and Importance Assignment.} At every iteration, our CCGR algorithm repairs a set of initial states and safeguards another. So far, our method selects these two regions based on sampled STL robustness and verification results. However, safeguarded simulated annealing is compatible with arbitrary regions to be repaired and protected, as long as we are able to compute STL robustness. Therefore, we can use arbitrary rules to decide what scenarios to repair and what to protect. Another flexibility is that we can assign different levels of importance to these initial state sets, by sampling different numbers of states or using multiple different balance factors in the energy function.

% \textbf{Using Verifier-Focused Objectives.} Right now, the CCGR algorithm improves an energy function based on STL robustness. This design assumes that the higher STL robustness in an initial state region, the more likely that this region will pass verification. However, this assumption is not necessarily true. As in the case studies, there are plenty of regions with high robustness but not passing Verisig, e.g., the green hatched regions in Figure \ref{fig:uuv_repair}. A future research direction is to study whether there exists a metric, on which verification results directly depend.

\textbf{Customized Criteria for Protection and Repair. }We use a sound but incomplete verifier to identify the states to be protected, and STL robustness to construct the objective function in \ourmethod{}. However, our algorithm can be flexibly applied to customized criteria. For example, one can use arbitrary rules like ``protecting the initial states in the living room" when repairing a service robot controller, simply because one cares about these initial conditions. User customization is also enabled by different numbers of sampled states per region $K$ and balance factor $\lambda$. Different $K$'s and $\lambda$'s can be assigned to different regions to signify degrees of importance.

\textbf{Connection to Preventing Catastrophic Forgetting.} As mentioned in Section \ref{sec:related_work}, the \ourproblem{} problem is similar to preventing catastrophic forgetting in continual and transfer learning: both aim to preserve existing knowledge in a learned agent.
% , but CCGR does not have an explicit target domain. 
A future research subject is to analyze the relationship between these two, to see whether the \ourproblem{} problem can be reduced to domain adaptation.

To summarize, this paper identified a gap in the repair algorithms of learning-enabled controllers, that current techniques have yet to preserve existing correct behaviors during repair. To fill this gap, we formulate the \ourproblem{} problem and its corresponding solution algorithm. Case studies show that our \ourmethod{} algorithm is not only able to preserve previously verified initial state regions --- but also repair a large set of incorrect ones, improving a CPS controller's STL robustness with respect to its task and outperforming simpler baselines.

\edit{\section*{Acknowledgement}}

\looseness=-1
\edit{This research was supported in part by NSF 2143274 and by ARO W911NF-20-1-0080. The views and conclusions contained in this document are those of the authors and should not be interpreted as representing the official policies, either expressed or implied, of the Army Research Office or the U.S. Government.}

% \bibliographystyle{plain}
% \bibliography{references}

\newpage
\appendix
\edit{\section*{Proof of Proposition 1}
\label{app:proof}

We hereby provide the proof of Proposition \ref{prop:bounded_error}.

\begin{proof}
% In Monte Carlo integration, 
Monte Carlo integration has the following property \cite{robert1999monte}: an integral $I = \int_{X} g(x) dx$ is approximated by $\hat{I} = (V/K)\sum_{x \in \hat{X}} g(x)$. Here, $\hat{X}$ consists of $K$ uniformly and independently sampled points in $X$ and volume $V = \int_{X}dx$. Existing work \cite{robert1999monte} has shown this approximation satisfies (i) $\mathbb{E}[\hat{I}] = I$ and (ii) $\text{Var}[\hat{I}] = V^2\text{Var}_{\hat{X}}[g]/K^2$.

The above is an established result of Monte Carlo integration. In our case, the energy function is a linear combination of two Monte Carlo integrals. Therefore, we have
\begin{equation}
    \label{eq:proof_1}
    \begin{split}
        &\mathbb{E}[\hat{e}(\theta)] = \frac{1}{K} \mathbb{E}[\sum_{s_0 \in \hat{S}_i} \rob(s_0, \pi_\theta)]\\
        &\quad\quad+ \frac{\lambda}{LK} \mathbb{E}[\sum_{s_0 \in \hat{S}^{\mathsf{s}}} \log(\rob(s_0, \pi_\theta))]\\
        &= \frac{1}{|S_i|} \underbrace{\mathbb{E}[\frac{|S_i|}{K}\sum_{s_0 \in \hat{S}_i} \rob(s_0, \pi_\theta)]}_{\mathbb{E}[\hat{I}]}\\
        &\quad\quad+ \frac{\lambda}{|S^{\mathsf{s}}|} \underbrace{\mathbb{E}[\frac{|S^{\mathsf{s}}|}{LK}\sum_{s_0 \in S^{\mathsf{s}}} \log(\rob(s_0, \pi_\theta))]}_{\mathbb{E}[\hat{I}]}\\
        &= \frac{1}{|S_i|}\int_{S_i} \rob(s_0, \pi_\theta) ds_0\\
        &\quad\quad+ \frac{\lambda}{|S^{\mathsf{s}}|}\int_{S^{\mathsf{s}}}\log(\rob(s_0, \pi_\theta)) ds_0 \\
        &= e(\theta).
    \end{split}
\end{equation}
Likewise, because the initial states are independently sampled,
\begin{equation}
    \label{eq:proof_2}
    \begin{split}
        &\text{Var}[\hat{e}(\theta)] =\frac{1}{K^2} \text{Var}[\sum_{s_0 \in \hat{S}_i} \rob(s_0, \pi_\theta)]\\
        &\quad\quad+ \frac{\lambda^2}{(LK)^2} \text{Var}[\sum_{s_0 \in \hat{S}^{\mathsf{s}}} \log(\rob(s_0, \pi_\theta))]\\
        &= \frac{1}{|S_i|^2} \underbrace{\text{Var}[\frac{|S_i|}{K}\sum_{s_0 \in \hat{S}_i} \rob(s_0, \pi_\theta)]}_{\text{Var}[\hat{I}]}\\
        &\quad\quad+ \frac{\lambda^2}{|S^{\mathsf{s}}|^2} \underbrace{\text{Var}[\frac{|S^{\mathsf{s}}|}{LK}\sum_{s_0 \in \hat{S}_2} \log(\rob(s_0, \pi_\theta))]}_{\text{Var}[\hat{I}]}\\
        &= \frac{1}{|S_i|^2}
        \frac{|S_i|^2\text{Var}_{\hat{S}_i}[\rob]}{K^2} + \frac{\lambda^2}{|S^{\mathsf{s}}|^2}\frac{|S^{\mathsf{s}}|^2\text{Var}_{\hat{S^{\mathsf{s}}}}[\log \circ \rob]}{(LK)^2}\\
        &=\frac{1}{K^2}\text{Var}_{\hat{S}_i}[\rob] + \frac{\lambda^2}{(LK)^2}\text{Var}_{\hat{S}^{\mathsf{s}}}[\log \circ \rob].
    \end{split}
\end{equation}
\end{proof}

\section*{Repeatability Package}
\label{app:rep}

% \subsection{Elements Included}
We provide a repeatability package that includes the code to reproduce the results of UUV and MC case studies as described in Section \ref{sec:case_studies}. It will output figures that distinguishes between three types of regions (counterexamples found, counterexamples not found but verification fails, verification succeeds) as in Figure \ref{fig:repair_results} using \texttt{matplotlib}. It will also output results in the same way as in Table \ref{tab:uuv_result} and Table \ref{tab:mc_result} in standard output. Moreover, running the verifier will output \texttt{.txt} log files that record verification time, and running the incremental repair algorithm will output repair time after each iteration. These are the times we recorded as in Table \ref{tab:uuv_time} and \ref{tab:mc_time}. Please refer to our \href{https://github.com/ericlupy/isar_rep}{GitHub repository} for detailed instructions.

The prerequisites of our code are already configured in the Docker image, and there is no specific requirement on the system. However, we recommend to run our code (especially the verification part) on a machine with a large number of CPUs available. This will significantly reduce the overall computational time by dispatching one thread per CPU. In our experiments, we parallelize the verification on 40 CPUs. }

\end{document}